\documentclass[12pt,dvips]{article}
\usepackage{amscd,verbatim}
\usepackage[all]{xy}
\usepackage{graphicx}

\usepackage{amsmath, amscd}
\usepackage[psamsfonts]{amssymb}
\usepackage{amsthm}
\usepackage{euscript}
\usepackage{amscd,verbatim}
\usepackage[all]{xy}
\usepackage{graphicx}

\usepackage{amsmath}

\usepackage{latexsym}

\renewcommand{\(}{\begin{equation}}
\renewcommand{\)}{end{equation} \vspace{-.05in}\linebreak}
\newcommand{\bea}{\begin{eqnarray}}
\newcommand{\eea}{\end{eqnarray}}

\newcounter{saveeqn}
\newcounter{savealpheqn}

\newcommand{\alpheqn}{\setcounter{saveeqn}{\value{equation}}%
      \stepcounter{saveeqn}\setcounter{equation}{0}%
      \renewcommand{\theequation}{\mbox{\arabic{section}.\arabic{saveeqn}
\alph{equation}}}
      \renewcommand{\)}{\end{equation}}}
\def\part#1{\frac{\partial}{\partial{#1}}}%
\def\group#1{\refstepcounter{equation}\setcounter{saveeqn}{\value{equati
on}}%
      \label{#1}\setcounter{equation}{0}%
\renewcommand{\theequation}{\mbox{\arabic{section}.\arabic{saveeqn}
\alph{equation}}}
      \renewcommand{\)}{\end{equation}}}
\newcommand{\reseteqn}{\setcounter{equation}{\value{saveeqn}}%
      \renewcommand{\theequation}{\arabic{section}.\arabic{equation}}%
      \renewcommand{\)}{\end{equation}}}

\newcommand{\aalpheqn}{\setcounter{saveeqn}{\value{equation}}%
      \stepcounter{saveeqn}\setcounter{equation}{0}%
      \renewcommand{\theequation}{\mbox{
            \Alph{subsection}.\arabic{saveeqn}\alph{equation}}}
       \renewcommand{\)}{\end{equation}}}
\newcommand{\areseteqn}{\setcounter{equation}{\value{saveeqn}}%
      \renewcommand{\theequation}{\Alph{subsection}.\arabic{equation}}%
      \renewcommand{\)}{\end{equation}}}

\renewcommand{\thefootnote}{\alph{footnote}}
\renewcommand{\(}{\begin{equation}}
\renewcommand{\)}{\end{equation}}
\newcommand{\ba}{\begin{eqnarray}}
\newcommand{\ea}{\end{eqnarray}}

\newcommand{\bp}{\mathop{\vtop{\ialign{##\crcr

$\hfil\displaystyle{}\hfil$\crcr\noalign{\kern-13pt\nointerlineskip}
       \BIG{(}\hskip0pt\crcr\noalign{\kern3pt}}}}}
\newcommand{\cbp}{\mathop{\vtop{\ialign{##\crcr

$\hfil\displaystyle{}\hfil$\crcr\noalign{\kern-13pt\nointerlineskip}
       \BIG{)}\hskip0pt\crcr\noalign{\kern3pt}}}}}
\newcommand{\pa}{\mathop{\vtop{\ialign{##\crcr

$\hfil\displaystyle{\oplus}\hfil$\crcr\noalign{\kern+1pt\nointerlineskip
}
       \hspace{.08in}$^{\alpha=0}$\hskip6pt\crcr\noalign{\kern3pt}}}}}

\newcommand{\R}{\ensuremath{\mathbb R}}

\newcommand{\C}{\ensuremath{\mathbb C}}

\newcommand{\Z}{\ensuremath{\mathbb Z}}

\def\dwn{\downarrow}

\newcommand{\beq}{\begin{equation}}
\newcommand{\eeq}{\end{equation}}

\newcommand{\iso}{\cong}

\numberwithin{equation}{section}

\catcode`\@=11
\def\vereq#1#2{\lower3pt\vbox{\baselineskip1.5pt \lineskip1.5pt
\ialign{$\m@th#1\hfill##\hfil$\crcr#2\crcr\sim\crcr}}}
\catcode`\@=12

\makeatletter
\newcommand\figcaption{\def\@captype{figure}\caption}
\newcommand\tabcaption{\def\@captype{table}\caption}
\makeatother
\renewcommand{\(}{\begin{equation}}
\renewcommand{\)}{\end{equation}}

\theoremstyle{plain}

\theoremstyle{definition}

\begin{document}

\begin{titlepage}
\begin{flushright}
hep-th/0701231 \\
\end{flushright}

\vspace{2em}
\def\thefootnote{\fnsymbol{footnote}}

\begin{center}
{\large\bf  The Loop Group of $E_8$ and Targets for Spacetime}
\end{center}
\vspace{1em}
\begin{center}
\Large Hisham Sati \footnote{E-mail:
hisham.sati@yale.edu}
\end{center}

\begin{center}
\vspace{1em}
{\em { Department of Mathematics\\
Yale University\\
New Haven, CT 06520\\
USA}}\\
\end{center}

\vspace{0.5cm}

\begin{abstract}
The dimensional reduction of the $E_8$ gauge theory in eleven dimensions 
leads to a loop bundle in ten dimensional type IA string theory. 
We show that the restriction to the Neveu-Schwarz sector leads naturally to a sigma 
model with target space $E_8$ with the ten-dimensional spacetime as the source.
The corresponding bundle has a structure group the group of based loops, whose 
classifying space we study. We explore some consequences of this proposal such as possible Lagrangians and existence of flat connections. 
\end{abstract}

\vfill

\end{titlepage}
\setcounter{footnote}{0}
\renewcommand{\thefootnote}{\arabic{footnote}}

\pagebreak

\renewcommand{\thepage}{\arabic{page}}

%\tableofcontents

%%%%%%%%%%%%%%%%%%%% %%%%%%%%%%%%%%%%%%%%%%%%%%%%%%%%%%%%%%%%%%%%%%%%
\section{Introduction}
%%%%%%%%%%%%%%%%%%%%%%%%%%%%%%%%%%%%%%%%%%%%%%%%%%%%%%%%%%%%%%%%%%%%

In the previous work \cite{S5} we studied the $E_8$ gauge theory over the circle part of the eleven-manifold, i.e. over the M-theory circle, making use of the Mickelsson construction \cite{Mick} of (usual) WZW model on $E_8$. The M-theory circle over $X^{10}$ gives rise to infinite-dimensional bundles. We consider the case of a product $Y^{11}=X^{10} \times S^1$. Then we have a loop group bundle $LE_8$
over $X^{10}$, essentially due to the fact that maps from a space $X$ to $LG$ is the same as
maps from $S^1 \times X$ to $G$. This gives a map $X^{10} \longrightarrow BLE_8$. Since the group of unbased loops $LE_8$ is topologically the product $\Omega E_8 \times E_8$ of the group of based loops and the finite dimensional Lie group, then we get two factors for the classifying space: $E_8$ and $BE_8$. 
The first is relevant for the description of the NS fields and the second for the RR fields 
(in particular the 4-form). We restrict to the NS fields and so we have a map 
$X^{10} \longrightarrow E_8$. 

\vspace{3mm}
The index of $E_8$ was discovered as part of the phase of the topological
part of the M-theory action \cite{Flux}. In \cite{DMW} the corresponding partition
function was related to the K-theoretic partition function of the Ramond-Ramond 
(RR) fields of type IIA string theory. The inclusion of the Neveu-Schwarz (NS) $H$-field
 was considered in \cite{MS}, where the description in terms of twisted K-theory 
 and the relation to loop group bundles was given, building on \cite{DMW} and on 
 the observation in \cite{AE}.  

\vspace{3mm}
In this note we study the restriction of the loop bundle to the based loops, meaning to 
the NS sector. This leads to several interesting consequences such as the appearance 
of a sigma model from ten-dimensional spacetime $X^{10}$ to $E_8$. 
We explore this from various angles, both physical and mathematical.
We emphasize that we give a modest proposal that does not address the 
big question: What is the physical nature of the $E_8$ gauge theory?
Instead we provide some observations that we hope will be useful for that
purpose. We also do not attempt to make any concrete constructions  
involving the description as a quantum field theory. While we do not 
answer such important questions, it is possible that eventually the putative `theory' 
we use will involve true topological field theory description in terms of path integrals or using 
some localization techniques. We hope that this approach will be useful in shedding some light
on such issues. Our use of the term "gauge theory" follows \cite{DMW}.

\vspace{3mm}
The note is organized as follows. In section {\bf \ref{trivial}} we study the restriction of 
the loops to the based loops, which means restricting to the NS sector in type IIA string theory.
We identify the classifying space for such bundles. This is where the sigma model 
emerges. 
In section {\bf \ref{terms}} we explore the possible terms 
in the action, namely the topological term and the chiral term. In section {\bf \ref{tft}} we motivate 
considering the $E_8$ gauge theory as a topological field theory along the lines of 
\cite{DFM} and study (a partial) analog of the construction of \cite{BJSV}. We also consider flat connections on 
the circle and their moduli space, which is just the circle itself via the holonomy. Next in 
section {\bf \ref{kk}} we revisit the Higgs field considered in \cite{S5} in the sigma model 
context. 

\vspace{3mm}
In {\bf \ref{higher}} we consider terms on $E_8$ other than the usual degree three
generator, which is, after all, is common to all Lie groups of dimensions greater or equal to three.
We explore whether something peculiar to $E_8$ can be found. At the rational level, 
wee could form the wedge products of the generators in different degrees. What stands out 
in this case is the degree three generator, wedged with degree fifteen generator giving a topological term in eighteen dimensions,
and with degree twenty-three generator to give a topological term in twenty-six dimensions. 
At the level of torsion (in cohomology) we could have several possibilities both in ten and higher dimensions. The interesting generators are related by cohomology operations and the connection to 
spacetime makes use of the fact that these operations commute with pullback. 
We comment on the nature of the map in section {\bf \ref{nature}} and consider consequences of 
the embedding and on the corresponding characteristic classes of spacetime. 
We conclude with further remarks on a mod 8 structure, a more general setting and global anomalies.
%%%%%%%%%%
\section{The Restriction to the Neveu-Schwarz Sector}
%%%%%%%%%%
\label{trivial}
We start with the bundle 
\(
%\left\{
\begin{matrix}
E_8&\to & P\cr
&&\dwn\cr
S^1& \to & Y\cr
&&\dwn \pi\cr
     & & X\cr
\end{matrix}
\)
and consider the product case $Y^{11}=X^{10} \times S^1$. In this case we 
have a loop bundle with structure group $LE_8$ over the base $X^{10}$ 
in type IIA string theory. This means that there is a map, the classifying map,
from $X^{10}$ to the classifying space $BLE_8$ of the $LE_8$ bundle. 
Since $LE_8$ is the semi-direct product $\Omega E_8 \ltimes E_8$, which we think 
of topologically a product, we ask whether applying the functor $B$, on the 
fibration
\(
\Omega E_8 \to LE_8 \to E_8,
\label{fib}
\)
i.e. taking the classifying the spaces corresponding to (\ref{fib}) would lead 
to splitting. The answer is negative as $BLG \neq B \Omega B \ltimes BG$. 
However, the restriction to $BLE_8 \supset B\Omega E_8$ can occur if the original 
$E_8$ bundle $P$ is trivial. In fact, there is a map from $X^{10}$
to $E_8$ if and only if $P$ is trivial, and in which case a trivialization
have to be fixed. In the rest of the paper we will assume this to hold.  
Before going any further we provide the argument for the above condition
on triviality of the bundle. Given the bundle $E$ with structure group $E_8$ over
$S^1 \times X^{10}$ we have a classifying map from $S^1 \times X^{10}$ to 
$BE_8$. Let us take $B$ of (\ref{fib})
\(
B\Omega E_8=E_8 \to BLE_8 \to BE_8.
\)
We are given a map $\gamma$ from $X^{10}$ to the middle factor $BLE_8$ 
($=LBE_8$ since $E_8$ is connected), and we ask to lift it to a map from $X^{10}$ to 
the first factor $E_8$, given that we have a restriction of $\xi : E \to S^1 \times X^{10}$
to a bundle on $X^{10}$, $\xi |_{X^{10}}$. If the bundle $\xi |_{X^{10}}$, corresponding to the third factor, is trivial then 
this implies that there is a bundle
$E'$ over $X^{10}$ with structure group $\Omega E_8$, corresponding to the 
first factor $\phi : X^{10} \to E_8$ , which in turn implies the existence of a bundle 
$E' \times_{\Omega E_8}LE_8$ corresponding to the second factor
$\gamma: X^{10} \to LBE_8=BLE_8$. This latter bundle is the pullback 
of the bundle with total space $LE_8$ (both with fibers $\Omega E_8$) from
$E_8$ to $X^{10}$, i.e. 
\(
\begin{matrix}
E' \times_{\Omega E_8} LE_8  & LE_8 \cr
\dwn &\dwn \cr
X^{10} & \to E_8,
\end{matrix}
\)
%\(
%\begin{matrix}
%E'&\buildrel{\overline{\phi}}\over{\to} & \bf{A}\cr
%\dwn&&\dwn\cr
%X^{10}& \buildrel{\phi}\over{\to} & E_8 
%\end{matrix}
%\)
both with fibers $\Omega E_8$, and where the second vertical arrow is just the 
bundle (\ref{fib}). We have the same fiber over spaces one of which is mapped 
to the other so we are tempted to ask whether this is a classifying map. Note however,
that the total space of the bundle over $E_8$ is not contractible and so cannot 
serve as a classifying space. Thus what we should aim for is a bundle with structure 
group $\Omega E_8$ over $E_8$ such that the total space is contractible.

\vspace{3mm}
The above discussion leads to the restriction of the bundle $E'$ with structure group 
$LE_8$ to a subbundle with structure group $\Omega E_8$, which was studied in 
in \cite{MS} and shown to represent the NS H-field. What is this principal $\Omega E_8$
bundle? Given the embedding $X^{10} \to E_8$, we ask the natural question whether 
bundles on $X^{10}$ can be viewed as pulled back from those over $E_8$.
The bundles in question are a priori infinite-dimensional and are the 
based loop groups $\Omega E_8$. In other words, can one view $E_8$ 
as the classifying space of such bundles, in the sense that the $\Omega E_8$
bundle on $X^{10}$ is obtained from that over $E_8$? We thus seek a space over
$E_8$ that would serve as a classifying space. In particular, this space has to 
be contractible by Kuiper's theorem.   

\vspace{3mm}
Let us look
at this from the point of view of connections $A$ on the circle which live in 
$\mathcal{A}$, the space of connections \cite{Mick}. The space of gauge orbits is 
$\mathcal{A}/{\mathcal{G}}$, where $\mathcal{G}$ is the group of based gauge 
transformations, i.e. the group of smooth maps $f: S^1 \to E_8$ such that 
$f(p)=1$ at a base point  $p\in S^1$. There is a universal bundle $P$ over
$E_8$ with total space the set of smooth paths $f:[0,1] \to E_8$ starting from the unit element 
and such that $f^{-1}df$ is periodic at the endpoints. 
What happens if we considered $LE_8$ bundles in place of $\Omega E_8$ bundles?
Consider the moduli space of $E_8$ bundles over $S^1$. This is $E_8$. For every point 
$q \in E_8$, there is a corresponding $E_8$ bundle $E_q \to S^1$ parametrized 
by $q$ and is determined up to isomorphism. A universal bundle \cite{AB} is an 
$E_8$ bundle $\mathcal{E} \to E_8 \times S^1$ such that for any $q \in E_8$, 
$\mathcal{E}$ restricted to $q \times S^1$ is isomorphic to $E_q$.
\footnote{The universal space can be seen by looking at the evaluation map 
$ev: S^1 \times {\rm Map}(S^1, BE_8) \to BE_8$, since a gauge theory is a sigma model 
with target $BG$ \cite{BCRS}. In our case we have an $E_8$ gauge theory on $S^1$.
Considering principal $G$-bundles
over $BE_8$ then one considers the commutative diagram obtained by pulling back 
bundles over $BE_8$ to bundles over $S^1 \times {\rm Map}(S^1, BE_8)$.}
Let us next look at the connections and the holonomies on $S^1 \times E_8$. 
The subset of smooth maps from $\R$ to $E_8$ given by
\(
\left\{  g(t): \R \to E_8 , g(0)=1, g(t+1)=g(t)\cdot g(1) \right\}
\)
is homeomorphic to the {\it topological affine Tits building} ${\bf A}(LE_8)$ \cite{Kit} (and 
the appendix in \cite{KM}).
The action of $h(t) \in LE_8$ on $g(t)$ is given by $hg(t)=h(t)\cdot g(t) \cdot h(0)^{-1}$.
The action of an element of the circle $x \in \R/\Z$ is given by 
$xg(t)=g(t+x) \cdot g(x)^{-1}$. The space  ${\bf A}(LE_8)$ is the desired model for $EH$ with $H$ the Kac-Moody
group $\widehat{LE_8}$, and  the two descriptions of  ${\bf A}$ are: the smooth 
infinite dimensional manifold of holonomies on $S^1 \times E_8$ and the affine space
$\mathcal{A}(S^1 \times E_8)$ on the trivial $E_8$ bundle $S^1 \times E_8$ \cite{Kit}.

\vspace{3mm}
To summarize, the $\Omega E_8$ bundles $E'$ over $X^{10}$ are classified by
$\bf{A}$, the space of $E_8$ connections on $S^1 \times E_8$. The latter is the total space 
of the trivial $E_8$ bundle on the circle, and the spaces fit into the diagram
\(
\begin{CD}
E' @>{\overline{\phi}}>> \bf{A} \\
@VVV  @VVV\\
X^{10} @>\phi>> E_8
\end{CD}
\)

with fibers $\Omega E_8$ -- so $\Omega E_8$ acts freely on the space of connections--
where the map from $\bf{A}$ to $E_8$ is the holonomy map hol.

%%%%%%%%%%%%%%%
\section{Terms in the Action}
%%%%%%%%%%%%%
\label{terms}
In this section we consider the possible terms in the action of the $E_8$ sigma model.  
Let us first look at the standard worldsheet case, i.e.  when the `domain' is a Riemann
surface $\Sigma$, and the corresponding map is $\phi: \Sigma \to E_8$. 
In this case one can have two terms:

\vspace{2mm}
\noindent {\bf 1. The chiral term}: Since $\pi_3(E_8)= \Z$ then the following term is possible
\(
S_{WZ}= \int_{\Sigma} Tr[ \phi^* \alpha \wedge * (\phi^* \alpha)],
\)
where $\phi^*(\alpha)$ is the one-form $g^{-1}(x) \partial_{\mu} g(x) dx^{\mu}$, and $\alpha$ is the Maurer-Cartan one-form.

\vspace{2mm}
\noindent {\bf 2. The B-field term}: One can also pull back the three form on the $E_8$ Lie group  
\footnote{One could use the basic gerbe on $E_8$ \cite{Mein}  to get the gerbe on the Handlebody
$\partial^{\dagger} \Sigma$ of $\Sigma$. }

\(
S_H=\int_{\Sigma} ``\phi^* B_2" = \int_{\partial^{\dagger}\Sigma} \phi^* H_3.
\)

 \vspace{3mm}
 Next, let us consider the ten-dimensional case. Here we have $\phi: X^{10} \to E_8$.
 We will look at the corresponding terms in the action: 

 \vspace{2mm}
 \noindent {\bf 1. The chiral term}: Since $\pi_{11}(E_8)$ is zero, there is no chiral term. However,
 if $E_8$ breaks into a subgroup $H \subset E_8$, then we could have $\pi_{11}(H)\neq 0$
and $\pi_{10}(H)=0$. This occurs for example for $E_7$, $E_6$, $F_4$ and others.
This can be easily seen since the homotopy types are well-known, and for the last two
Lie groups, for instance, are given by $(3,9,11,15,17,23)$ and $(3,11,15,23)$, respectively. 
In this case, up to a normalization, the chiral term is
\(
S_{WZ}=\int_{X^{10}} {\rm Tr}\left(  \underbrace{\phi^* \alpha \wedge \phi^* \alpha
 \wedge \cdots \wedge \phi^* \alpha}_{10} \right).
\)

\vspace{2mm}
  \noindent {\bf 2. The B-field term}: Again here the sparsity of the homotopy groups of $E_8$ makes
  a difference. Since in the range from $0$ to $10$ the only nontrivial homotopy group 
  is $\pi_3$, then we can only pull back the three-form as in the case for the Riemann 
  surface. In order to get a ten form, we write the action as (again up to proper normalization) 
  \(
  S_H=\int_{\partial^{\dagger} X^{10}} \phi^* H_3 \wedge Z_8.
  \label{Z8}
  \) 
  Here $Z_8$ is viewed as an auxiliary eight-form as far as $E_8$ is concerned. The two
  obvious examples of this are $F_4 \wedge F_4$, coming from the Chern-Simons term,
  and $Z_8=I_8$, the Green-Schwarz polynomial coming from the one-loop term. 
 $Z_8$ is closed in the latter and is closed in the former provided $F_2^{RR}$, the RR
 field of the circle, is zero. In this note we are restricting to the Neveu-Schwarz sector. 
 We will look further at the possibilities in section {\bf \ref{higher}}.

\vspace{3mm} 
Interestingly, this makes use of the coboundary `theory' proposed in \cite{S5}. The existence
has implications on $X^{10}$, one of which being that all its Stiefel-Whitney numbers 
 vanish.
 
 %%%%%%%%%%%
\section{`Topological Field Theory' and Flat Connections?}
%%%%%%%%%%
\label{tft}

An obvious question is: What is the nature of the $E_8$ gauge theory? While we do not
have complete and precise answers, we give arguments that we hope provide some hints
and possibilities. 
The intuition is that 
this $E_8$ gauge 
theory is some kind of topological field theory in the bulk of the eleven-dimensional spacetime.
One can give the following arguments for this \cite{DFM}. 
One can shift the connection $A$ on the principal bundle $Q$, with 
characteristic class $a$, to any other connection $A'$ in the space $\mathcal{A}(Q(a))$
of smooth connections and hence $A$ is only constrained by topology. 
The relation to the heterotic boundary
is that this theory becomes dynamical on the spatial boundary. In the bulk, the term 
$\int_{Y^{11}} {\rm Tr}  F \wedge * F$ is not gauge invariant and so there are no 
propagating degrees of freedom on $Y^{11}$. However, such a term becomes 
gauge invariant, and hence dynamical, over the ten-dimensional boundary \cite{DFM}, 
where the notion of groupoid of fields is used.

\vspace{3mm}
Next, we discuss some possibilities resulting in the topological nature of this theory. 
The first questions is whether the theory is supersymmetric. This has been considered
in \cite{ES} from a different point of view. We will leave the possibility that certain 
supersymmetric extension may exist. 
\footnote{An extension for the pure Yang-Mills case in eleven dimensions is given is 
\cite{Nish} with a Lorentz-invariant Lagrangian whose symmetry is broken spontaneously
i.e. at the level of field equations.} 
Pure Yang-Mills is of course not topological but becomes so by adding the right
amount of supersymmetry and then making the appropriate twists to define new kinds
of fields from old ones. It is possible that twists might make the $E_8$ gauge theory topological 
as happens for Yang-Mills theory.
\footnote{An obvious  caveat here is that a usual twist requires an R-symmetry, 
something that does not exist if the theory has $\mathcal{N}=1$ supersymmetry. 
For a discussion of topological field theories in higher dimensions in relation to 
supersymmetry see \cite{BL}. }  
At any rate, this desirable analog in the case of the $E_8$ gauge theory
at hand seems to be a nontrivial task and is beyond the scope of this note. 

\vspace{3mm}
The alternative, i.e. that the theory is manifestly metric independent, i.e. 
topological without kinetic terms for the gauge fields, would lead to flat connections 
in a natural way since the connections related only to topology are the flat ones. 
The Chern-Simons-like nature of the $C$-field would also give
flat connections because those are the critical points of the Chern-Simons action
( taken as that of the membrane). The $C$-field is not quite $CS(A)$ on the nose 
but a modification of it. Neither is it a three-form $c \in \Omega^3(Y^{11})$. One model 
for the $C$-field in 
\cite{DFM} gives the $C$-field as a pair $(A, c)$ such that the fourth differential
cohomology group, twisted by half the string class $\frac{1}{2}\lambda$, is    
\(
\check{H}_{\frac{1}{2}\lambda}^4 (Y^{11}) = \mathcal{A} (Q(a)) \times \Omega^3(Y^{11})
/ \Omega^1({\rm ad}Q) \ltimes \Omega_{\Z}^3(Y^{11}).
\)
In any case, there is still a Chern-Simons `part' to $C$, and, when taken as an action,
will have an equation of motion given by flat connections $F=0$.

%%%%%%%%%%%%%
\subsection{The Generalized Sigma Model via the Adiabatic Limit }
%%%%%%%%%%%%
We have, as in \cite{MS},  the eleven-manifold $Y^{11}$ with 
metric $g_{Y^{11}}=g_{S^1} + g_{X^{10}}$, such that the 
connection one-form splits as $A = A_{S^1} + A_{X^{10}}$. The analysis in \cite{MS}
and the interpretation there and in \cite{S5} gives the scaling upward of the metric 
on the base to large volume as corresponding to the adiabatic limit. Alternatively one 
could look at the small volume limit of the fiber, i.e. scaling downward the metric 
on the fiber $g_{S^1} \to t g_{S^1}$. The Yang-Mills action $\int_{Y^{11}} {\rm Tr} F \wedge *F$
splits into components in a similar way as happens for (supersymmetric) Yang-Mills 
theory in four dimensions \cite{BJSV}. However, unlike \cite{BJSV}, we do not get 
a piece that is completely over $S^1$ simply because $F_{\theta \theta}=0$, where $\theta$ is the 
circle direction. In any case, connections on the circle are always flat (we will come back to this soon).

\vspace{3mm}
In order to get a component of $F$ on the fiber we could work with the disk bundle
$\mathbb{D}^2 \to Z^{12} \to X^{10}$, where $Z^{12}$ is the theory whose boundary gives
M-theory. Indeed, the analysis of Witten \cite{Flux} leading to the appearance of $E_8$ started with
this theory, so we know that the $E_8$ bundle extends to $Z^{12}$. We now perform the
scaling down to the metric of the fiber given by the disk $\mathbb{D}^2$. In this case
we have a splitting analogous to that of \cite{BJSV} and the Yang-Mills kinetic term
$\int_{Z^{12}} {\rm Tr} F \wedge *F$ gives $ {\rm Tr}  F_{\mathbb{D}^2} 
\wedge * F_{\mathbb{D}^2}$, scaled with negative powers of $t$ and terms of order 
$\mathcal{O}(t^{\geq 0})$.
Requiring finiteness of the action in the limit $t \to 0$ imposes the flatness condition 
on the fiber components $F_{\mathbb{D}^2}=0$. 
\footnote{See e.g. \cite{disk} for some analytical aspects of connections for the case of boundary
including that of the disk.}

\vspace{3mm}
Here we go back to the flat connections on the circle. Since $\pi_1(S^1)=\Z \neq 0$, the flat connections
on $S^1$ are not trivial, since they can have a non-zero holonomy, for instance given by 
a Wilson loop, around a nontrivial loop on $S^1$. In fact all loops are nontrivial. The holonomy of each flat connection is an element 
of $E_8$ parametrized by a loop around $S^1$, and so defines a representation of $\pi_1(S^1)$ 
in $E_8$. Such representations are given by 
\footnote{$\pi_1(S^1)$ can be viewed as the classifying space $BS^1$.}
${\rm Hom}(\pi_1(S^1), E_8)$. 

\vspace{3mm}
As in \cite{BJSV}, specifying a flat connection $A_{S^1}$ on $X^{10} \times S^1$ 
amounts to specifying a map $\phi: X^{10} \to \mathcal{M}(S^1)=E_8$. The dependence
of the connection $A_{S^1}(\theta, x^i)$ on $\theta$ and on $x^i$, the coordinates
on spacetime $X^{10}$, then gets modified to $A_{S^1} (\theta, \phi(x^i))$,   
i.e. to depend on $\theta$ and the embedding $\phi(x^i)$. 
The flatness condition implies the nilpotency of the covariant derivative $D_{S^1}^2=0$,
so that the tangent space to the moduli space is given by $D_{S^1}$-cohomology 
$H^1(S^1, \frak{e}_8)$. 
What is $A_{X^{10}}$? As in \cite{BJSV} $A_{X^{10}}$
plays the role of an auxiliary field since the action is quadratic in $A_{X^{10}}$
and does not depend on $\partial_{X^{10}} A_{X^{10}}$. 
The variation of the flat connection $\delta A_{S^1}$ can be decomposed with respect
to a basis $\{ \alpha_I\} \subset H^1(S^1, \frak{e}_8)$ modulo gauge transformations
\(
\frac{\partial A_{S^1}}{\partial y^I}= \alpha_I + D_{S^1} E_I,
\)
where $E_I$ defines the connection on $\mathcal{M}(S^1)=E_8$. 

\vspace{3mm}
The Laplacian $D_{S^1}D_{S^1}$ can be inverted if the connection on $S^1$ is irreducible.
This happens when the holonomy group is precisely $E_8$ and not a proper subgroup, and 
implies that every group element in $E_8$ is realizable as a parallel transport. A reducible flat connection would give rise to an abelian representation
of $\pi_1(S^1)$ in an abelian subgroup of $ E_8$. However, since  flat connections are in one-to-one
correspondence with conjugacy classes of representations of $\pi_1(S^1)$ in $E_8$, and these
classes are abelian, then we cannot have irreducible connections for $E_8$. However, if 
$E_8$ is broken down to $U(1)$ then we can have irreducible connections.
In this case the Laplacian  can be inverted to get for the connection
$A_{X^{10}}=E_I \partial_{X^{10}} Y^I$.
Substituting into the action one gets a ten-dimensional sigma-model action
\(
S= \int_{X^{10}}  h_{IJ} \partial_{x^i}Y^I \wedge *(\partial_{x^j} Y^J) g_{ij},
\)
where $g_{ij}$ is the metric on $X^{10}$ and $h_{IJ}$ is the metric on $E_8$. 

\vspace{3mm}
Depending on structures on $X^{10}$ one can have variations on the construction. 
For example, if $X^{10}$ is complex then one could use complex coordinates, and 
so on.

%%%%%%%%%%%%%%%
\subsection{Moduli space of $E_8$ bundles on the circle}
%%%%%%%%%%%%%%
\label{moduli}

%%%%%%%%%%%%%%%%%
Consider the principal $E_8$ bundle $\pi$ :  $P \to S^1$
on the circle with a marked point 
$p_0 \in P$ corresponding to the point $\pi(p_0)=0$ on $S^1$.
The isomorphism classes ${\rm Bun}_{E_8}(S^1)$ of such bundles are classified by
their holonomy, and are given by homotopy classes of maps from $S^1$ to 
the classifying space $BE_8$. This in turn is given by the fundamental group
of $BE_8$ which is the same as the group $E_8$ itself. 

\vspace{3mm}
The fact that $\mathcal{M}(S^1)$ is equal to $E_8$ can be more intuitively seen for 
the case of a finite group in place of $E_8$. Let us take $G=\Z_2$, then the possible 
principal $\Z_2$-bundles over $S^1$ correspond to coverings. The trivial cover
corresponds to a product of two circles over our base circle, and there is one
more cover given by a helix with two loops and with the endpoints joined together.
The latter is the nontrivial $\Z_2$ bundle. We can easily see that these are the only
two, and so in this case $\mathcal{M}(S^1)$ is just the set $\Z_2$.

\vspace{3mm}
Let us give some description of the holonomy. The completion of one rotation around the
circle can be considered by looking at an interval $[0,1]$ and looking at the value
of the lift $\tilde{s}: [0,1] \to P$ of the exponential map $s: [0,1] \to S^1$, which takes
$0\leq t \leq1$ to $\exp(2\pi i t)$ in such a way that $s=\pi \circ {\tilde{s}}$. The images
of $\tilde{s}(0)$ and $\tilde{s}(1)$ belong to the same fiber $\pi^{-1}(0)$ over the marked point
$0$ of $S^1$,  which means that they can only differ by the action of an element $g$ of $E_8$,
i.e. $\tilde{s}(0)=\tilde{s}(1) \cdot g$. 
What effect does the fact that ${\rm Bun}_{E_8}(S^1)=E_8$ has? One effect is the following.
In general there is a natural action of $G$ on ${\rm Bun}_G$. In our case, there is thus an action 
of $E_8$ on itself under the holonomy isomorphism ${\rm hol}:  {\rm Bun}_{E_8}(S^1) \to E_8$
given by the conjugation action of $E_8$ on itself.

\vspace{3mm}
The space of equivalence classes of principal bundles can be determined using 
the fundamental group. Fix a basepoint $m \in S^1$. Over this point, the fiber is 
$P_m$. The bundle $P$ determines a map from the loops in $S^1$ based at 
$m$ to $E_8$ by taking the holonomy around the loop using the basepoint $p$
in $P$ corresponding to $m$. The loop around the circle lifts in the fiber to a loop
that starts at $p$ and ends at a point given by $p'=p\cdot  h$ where $h \in E_8$ 
is the holonomy that only depends on the homotopy class of the loop. 

\vspace{3mm}
Following \cite{F95} the category 
\footnote{The objects are elements in $G$ and the morphisms are conjugation actions.}
of principal $E_8$ bundles over $S^1$ 
with a chosen basepoint covering the basepoint in $S^1$
can be described. The holonomy around $S^1$ describes a map 
\footnote{The prime indicates fixing a basepoint, and the overline 
indicates isomorphism classes.} ${\rm hol}: {\mathcal{C}}'(S^1) \to E_8$,
that is an isomorphism on equivalence classes $\overline{\mathcal{C}}(S^1) \iso E_8$.
The conjugacy classes of the holonomy $g^{-1}hg$ arise from the change of the holonomy 
due to the change of basepoint as $p \to p \cdot g$ and are independent of the basepoint,
so that the isomorphism classes of principal bundles over the circle are given \cite{F95}:
${\overline{\mathcal{C}}}(S^1) \iso {\rm conjugacy~classes~in~} E_8$.

\vspace{3mm}
In \cite{BV} this issue of the basepoint for the circle pulled back from the base $X^{10}$ 
which is identified with the M-theory circle led to the inclusion of this circle as $U(1)_{\rm rot}$,
corresponding to the rotation of the loops in the case of nontrivial circle bundle.

%%%%%%%%%%%%%%%%%%%%%%%
\section{The Kaluza-Klein Assumption and the Higgs Field}
%%%%%%%%%%%%%%%%%%%%%%%%
\label{kk}
Here we consider the case where there is no dependence on the vertical direction(s).
 Let $A$ be a connection on the principal $E_8$ bundle $P$ over $Y^{11}$ and 
 $F(A)$ its curvature. Associated to $P$, via the adjoint representation of $E_8$ on 
 its Lie algebra $\frak{e}_8$, is the adjoint vector bundle 
 ${\rm ad} P= P \times_{E_8} \frak{e}_8$. The curvature is a two-form with values in 
 the adjoint bundle $F(A) \in \Omega^2(Y^{11}; {\rm ad} P)$. Locally, the components
 are  $F(A)=\sum_{M<N}^{11} F_{MN} dx^M \wedge dx^N$. With respect to a trivialization,
 the connection is described by a Lie algebra-valued one-form
  $A= \sum_{M=1}^{11} A_M dx^M$. The curvature can be expressed in terms of the 
  connection as $F(A)=dA + A^2$ or, using the covariant derivative 
  $\nabla_M = \partial_M + A_M$, as $F_{MN}=[\nabla_M, \nabla_N]$.  
  
\vspace{3mm}
The Kaluza-Klein assumption is that the Lie algebra-valued functions $A_M$ are 
independent of the eleventh coordinate, $\theta$. This means that these connections
define functions $A_{\mu}$ of the ten basic coordinates $x_{\mu} \in X^{10}$ ($\mu=0, \cdots, 10$)
giving the connection $A= \sum_{\mu=1}^{10} A_{\mu} dx^{\mu}$ over $X^{10}$. 
The eleventh component $A_{11}$ is an auxilliary field over $X^{10}$ which is the Higgs field
$\phi$. In coordinate-free language, the KK assumption is equivalent to requiring invariance
under the action, via $a$, of the circle group $S^1$ as $a\cdot(x_{\mu}, \theta)=(x_{\mu}, \theta +a)$.  
In this context, the Higgs field is defined as the difference of the covariant derivative and the 
Lie derivative $\phi= \nabla_{\theta} - \mathcal{L}_{\partial_{\theta}}$.

\vspace{3mm}
The principal bundle $P$ over $Y^{11}$ restricts to a principal bundle $P$ (we use the same notation)
over $X^{10}$ with the induced connection. Over the base, the fields are then 
$F \in \Omega^2(X^{10}; {\rm ad} P)$ and $\phi \in \Omega^0(X^{10}; {\rm ad} P)$.  
In our case the flatness condition $F(A)=0$ does not come from self-duality as in 
\cite{Hit} but rather from the requirement of Fourier decomposition \cite{S5}. The condition is
the homotopy-flatness of the connection \cite{CS}. If we restrict the connections to be
actually flat, then, in our ten-dimensional case, we have
\bea
F_{\mu \nu}&=&[\nabla_{\mu}, \nabla_{\nu}]=0
\nonumber\\
F_{\mu \theta}&=&[\nabla_{\mu}, \phi ]=0,
\eea
so the Higgs field is a covariantly constant function on $X^{10}$. This corresponds locally
to harmonic maps \cite{Hit} from $X^{10}$ to $E_8$.  Here again we get the map from spacetime
to $E_8$.  One consequence of the above is that the string class, given by \cite{Hig}
\(
-\frac{1}{4\pi^2} \int_{S^1} \langle F, \nabla \phi \rangle d\theta,
\)
would vanish. 
\footnote{In \cite{Hig}, the covariant derivative involves $\partial_{\theta}$ but this does not 
affect the conclusion due to the Kaluza-Klein assumption.}
Thus the flatness condition 
is equivalent to the vanishing of the $H$-field.

%%%%%%%%%%%%
\section{Higher Degree Generators}
%%%%%%%%%%%%
\label{higher}
Since we took the step to replace the worldsheet in the usual WZW model with the 
ten-dimensional spacetime-- which traditionally acts as the target for the worldsheet,
but now acts as a `domain' of an $E_8$ sigma model -- we ask whether there is 
anything interesting beyond dimension ten . We will see that indeed there is. After all, 
there does not seem
to be anything particularly special about dimension ten as far as $E_8$ is concerned (at least not yet). 
Since a sigma model involves pulling back natural forms from the target, the most interesting 
dimensions will be those which correspond to natural generators on $E_8$.  We consider this next,
first rationally and then including torsion. 

\subsection{Rationally}

Since the homotopy
type of $E_8$ is $3,15,23,27, \cdots$ we can form the wedge product of two generators
of different degrees to form an action over spaces of dimensions not restricted to 
the dimensions of the rational generators. We see that the next two interesting dimensions
to replace the auxiliary eight-form of ten dimensions with something more intrinsic to 
$E_8$ is dimensions eighteen and twenty six.  
We can form the topological terms $\int_{X^{18}}H_3 \wedge H_{15}$ and 
$\int_{X^{26}}H_3 \wedge H_{23}$ in these dimensions,  
respectively. 
Here we pull back the generators $\omega_3$, $\omega_{15}$, $\omega_{23}$ of the 
cohomology of $E_8$ via the map $\phi$ from spacetime $H_3 = \phi^* \omega_3$,
$H_{15}=\phi^* \omega_{15}$, and $H_{23}=\phi^* \omega_{15}$. 
This is an interesting way in which $E_8$ knows about the special dimensions 
of string theories.

%%%%%
\subsection{Torsion}
%%%%%%%%%%%%%%%%%%%%%%%%%%%%%%%%%%%%%%%%%%%%%%%%%%%%%%%%
\label{tor}
The H-field can be seen to represent the degree three class of 
an $\Omega E_8$ obtained by dimensional reduction on $S^1$ from the 
$E_8$ bundle in eleven dimensions \cite{AE,MS}.
The classifying space for $\Omega E_8$ bundle is $B \Omega E_8$,
which is just $E_8$ itself. Thus, in order to consider torsion classes 
of $\Omega E_8$ bundles we have to look at cohomology of $E_8$ at 
these primes. In the range of dimensions of interest to us, namely
less or equal to ten, the cohomology groups (more precisely, rings) 
are as follows. For ${p=2}$
\(
H^*(E_8;\Z_2)=\Z_2[x_3,x_5,x_9]/\left( x_3^{16}, x_5^8, x_9^4 \right),
\label{2}
\) 
where the generators $x_i$ are primitive and are related via the 
Steenrod squares in the mod 2 Steenrod algebra 
$Sq^2 x_3 = x_5$ and $Sq^4 Sq^2 x_3 = x_9$.
For ${p=3}$
\(
  H^*(E_8;\Z_3)=\Z_3[x_8]/(x_8^3)
\otimes \Lambda(x_3,x_7),
\label{3}
\)
where the generators $x_i$ are again primitive in our range of 
dimensions and are related via the Steenrod powers in the mod 3 
Steenrod algebra as $P^1 x_3 = x_7$ and $\beta P^1 x_3 = x_8$.
For $p=5$ we only have the generator $x_3$ in our range of dimensions. For
${p>5}$,  
\(
H^*(E_8;\Z_{p>5})=\Lambda(x_3),
\label{rat}
\)
with a primitive $x_3$. 

\vspace{3mm}
From (\ref{rat}) we see that the torsion in cohomology away from the torsion primes,
$p=2,3$ and 5, of $E_8$ is very simple and is the same as one would 
obtain in the rational case. The interesting torsion can be easily seen 
from the above equations (\ref{2}), (\ref{3}) to be encoded in the 
$E_8$ torsion primes. 
How are the above related to the cohomology classes on $X^{10}$? 
The answer is that this relation is given by the homomorphism induced
by the mapping $ \phi: X^{10} \to E_8$. This is possible because 
the Steenrod squares and the Steenrod powers commute with pullback, $\phi^*$,
\bea
Sq^s \phi^* &=& \phi^* Sq^s
\nonumber\\
P^s \phi^* &=& \phi^* P^s.
\eea
This implies that the pullback of the generators $x_i$ of the cohomology ring of $E_8$ at the 
various primes is the same as the application of the power operations on the the pullback
of the degree three class, which is essentially the $H$-field. In addition to the degree three
generator we can have degree five $x_5$ and nine $x_9$ generators, and degree
seven and eight generators $x_7$ and $x_8$, obtained 
from $x_3$ by the application of the cohomology operations as above. This shows that 
torsion gives further possibilities than the rational case.

%%%%%%%%%
 \section{Nature of the Map}
 %%%%%%%%%%
 \label{nature}
 We have seen that the loop group description of the NS H-field leads to a map
 from $X^{10}$ to $E_8$. What is the nature of such a map? We will consider two
 aspects of this question next. While this will not be essential for the physics, i.e.
 for pulling back the fields or defining the path integral, we include it to illustrate 
 some concrete situations.
  
 %%%%%%%
 \subsection{The map as an embedding}
 %%%%%%%%
 In order to deal with pullbacks one usually requires the given map to be an embedding.
\footnote{However, for defining a path integral this is not essential.}
 The question of whether a given map is an embedding is in general a difficult questions
 and it seems to have answers only in special cases. However, the map we have is 
 a classifying map which means that it is given up to homotopy. By classic consideration in
 differential topology, a map between two space $X^n$ and $Y^m$ is homotopic to 
 a differentiable embedding if we are in the so-called stable range, i.e. if $m .\geq 2n+1$.
 The intuition is that in this case there is enough room -- i.e. codimension-- to 
 resolve any tranversality problems. For $n=10$ and $Y^m =E_8$ we are well within 
 the stable range and so our map is homotopic to a differentiable embedding.
 One can then ask whether the ``homotopic" part of the statement can be 
 removed. In any case it might even
 be enough to have the map to be just homotopic to a differentiable embedding because 
 in that case we can still pull back closed forms. However, if  we insist to have an embedding
 then we are guaranteed to have one (in fact many) within the homotopy class. 
 \footnote{I thank Andrew Casson for a discussion on this point.}

 \vspace{3mm}
 There are consequences for the map to be a differentiable embedding. If we assume that 
 a map cobordant to a differentiable embedding is in fact a differentiable embedding then
 such conditions involve the mod 2 Stiefel-Whitney classes ( see \cite{KY}).  
 Applied to our situation, they are that $w_i(\phi)=0$ for 
 $i > k={\rm dim}(E_8) - {\rm dim}(X^{10})=238$, and $\phi^*\phi_{!}(1)= w_{238}(\phi)$.
Here the Stiefel-Whitney class $w_i(\phi)$ is defined as the cup product $\overline{w}(X^{10})
\cup \phi^* w(E_8)$. Since all characteristic classes, including the Stiefel-Whitney classes, 
of Lie groups are zero, then the first condition is satisfied and the second condition is  
$\phi^*\phi_{!}(1)=0$.  

\vspace{3mm}
Note that if we knew that the map was a topological embedding and 
asked for whether it is a differentiable embedding then the task would have been much easier.
Classic results (see \cite{Hae}) say that a topological embedding, $X^{n}$ in $M^m$, is 
a differentiable one if $m \geq \frac{3(n+1)}{2}$. In our case, this is obviously satisfied as we 
are well within the range. This discussion matters more when we break the group $E_8$ down
to a subgroup $H$ and start lowering the codimension of $X^{10}$ in $H$. For example, 
in the breaking to unitary groups, once we reach $SU(4)$ we enter the unstable range and 
we are forced to take the obstructions into account.

%%%%%%%%%%%%%%%%%%%%%%%%%%%%
\subsection{Some consequences for characteristic classes}
%%%%%%%%%%%%%%%%%%%%%%%%%%%%%
In the two-dimensional case, only the first Chern class could take on nonzero values. However in
the ten-dimensional case, the higher classes can in principle be nontrivial and so it is useful to give
some characterization of these. 

\vspace{3mm}
From the embedding of $X^{10}$ in $E_8$, 
we get the short exact sequence on the corresponding tangent  bundles
\(
0 \to TX^{10} \to TE_8 \vert_{X^{10}} \to N \to 0,
\)
where $N$ is the normal bundle to the embedding. First note that $TE_8$ is trivial 
as a vector bundle. This is because Lie groups are parallelizable. This can be easily seen as
follows. The Lie algebra $\frak{e}_8$ can be viewed as $T_eE_8$, and the map
from $ E_8 \times \frak{e}_8$ to $T_eE_8$ taking $(g, X)$ to $(g, (dL_g)_eX)$ is an isomorphism 
of vector bundles, i.e. $TE_8$ is the product vector bundle. This implies that all 
characteristic classes of $TE_8$ are trivial. 
The Whitney product formula
for the Pontrjagin classes gives (see e.g. \cite{BT})
\(
p(TE_8\vert_{X^{10}} )= p(TX^{10}) p (N).
\)
Since the left hand side is trivial, this means that any non-triviality in the classes 
of $TX^{10}$ will come from requiring the normal bundle to be nontrivial. Since the 
restriction $TE_8\vert_{X^{10}}$ is the pullback of $TE_8$ under the embedding 
$\phi : X^{10} \to E_8$, it follows from the functoriality of the total Pontrjagin class
that $p(TE_8\vert_{X^{10}})=\phi^*p(TE_8)=1$. This implies that 
\(
p(N)=\frac{1}{p(TX^{10})}.
\)

\vspace{3mm}
Given a particular manifold $X^{10}$ with known Pontrjagin classes, the corresponding classes
of the normal bundle can be determined. Alternatively, given certain conditions on the embedding, 
these give values to $p(N)$, which in turn constraint the allowed values of $p(TX^{10})$, and hence
the allowed manifolds $X^{10}$.  If $N$ is trivial then, in particular, the condition $p_1(TX^{10})=0$ for having a string structure will follow from triviality of $TE_8$. 
Similar analysis follows for the other classes,  for instance the 
Chern classes $c_i$ and the Todd classes $Td_i$ if we consider the complexification.

%%%%%%%%
\section{Further Remarks and Discussion}
%%%%%%%
The group $Ad E_8$ ($= E_8$) acts on a representation by conjugation to give an 
equivalent one, so that the set of flat connections is the quotient
${\rm Hom}(\pi_1(S^1), E_8)/Ad E_8$,
which is just `$E_8/E_8$'. In this context, the basic $E_8$-equivariant gerbe can 
be constructed as in \cite{Mein}. It would be interesting to study how far the analogy with 
the $G/G$ model can be carried. The fact that the isomorphism classes of principal bundles
over the circle are given by the conjugacy classes draws some resemblance to the case
of D-branes in WZW models, and suggests incorporating the boundaries $\partial X^{10}$ 
this way.  One can also ask whether other groups can occur as targets. 
Indeed, this is simply the reduction of the structure group $E_8$ to the unitary groups along 
the lines of \cite{DMW}, or, in the infinite dimensional setting from the type IIA point of view, via 
the breaking patterns discussed in \cite{S5}. Another direction of generalization is the 
inclusion of the $E_8$ part of $LE_8$. Recall that we have restricted to the subgroup $\Omega E_8$
of based loops, whose classifying space if $E_8$ itself. Including the `complement' $E_8$ of 
$\Omega E_8$ will lead to maps to the classifying space $BE_8$. This is just another way
of looking at usual gauge theory, i.e. as a sigma model with target space $BE_8$ \cite{BCRS}. 
What we get from this is, for instance, the generator of degree four $x_4$ coming from $H^*(BE_8)$ which complements the discussion in ${\bf \ref{tor}}$.

\vspace{3mm}
 {\bf 1. {More general setting}:}
 We have used the loop group description to give a sigma model with target the Lie group.
One could ask whether more general settings can be considered. In one direction of generalization, 
this would 
involve the space ${\rm Map}(X^{10}, G)$. We expect such situations to occur and we believe
they are interesting to study. For example, when $X^{10}=\R^{10-m} \times T^m$, 
for $m=2, \cdots, 10$,
then toroidal generalizations of the loop group make an appearance. Another point is that 
our discussion has been mostly topological. One might eventually be interested in adding structures, for instance holomorphic or K\"ahler. Further, quantum considerations are important 
and should be taken into account. In the WZW context, steps in this direction have been taken in \cite{avatars} and \cite{IKU}.

\vspace{3mm}
%%%%%%%%%
{\bf 2. {Mod 8 periodicity of structure}:}
%%%%%%%%%
The observations in this note and in \cite{S5} imply similarities 
 between worldvolume theories in the usual WZW 
 constructions and spacetime theories. This suggests the mod 8 periodicity  
 as indicated in the table:
\begin{center}
\begin{tabular}{lll}
 & Dimension &  \hspace{0.5cm} Dimension + 8\\
\hline
``Spacetime" & \hspace{0.5cm} $\int B_2$ & \hspace{1cm}  $\int B_2 \wedge Z_8$\\
\hline
Circle Bundle & \hspace{0.5cm} $\int C_3$ & \hspace{1cm}  $\int C_3 \wedge Z_8$\\
\hline
 Handlebody & \hspace{0.5cm} $\int H_3$ &  \hspace{1cm}  $\int H_3 \wedge Z_8$\\
\hline
Disk Bundle & \hspace{0.5cm}  $\int F_4$ &  \hspace{1cm} $\int F_4 \wedge Z_8$\\ 
\hline 
\end{tabular}
\end{center}
The second and third rows correspond, respectively, to going though the circle bundle first 
then taking the coboundary or to taking the coboundary 
first and then the circle bundle, arriving through both paths to the disk bundle, represented by the fourth row. 

%%%%%%
\vspace{3mm}
{\bf 3. {No global anomalies}:}
%%%%%%
A natural question is whether the generalized sigma model with target $E_8$ is free of global anomalies. First, for the $E_8$ gauge theory on the eleven-dimensional space $Y^{11}$,
the condition is that the cohomology group $H^3(Y^{11}, \Z)$ is torsionless. 
For the sigma model, applied to our situation, the condition is the vanishing of the 
cohomology group $H^{12}(E_8,\Z)$
\cite{MN}, or more precisely the vanishing of the torsion part 
${\rm Tor}H^{12}(E_8, \Z)={\rm Tor}H_{11}(E_8, \Z)$ \cite{BCRS}. All of
this group vanishes for $E_8$, and so the model does not suffer from global anomalies. 
However, again, the breaking to subgroups may change this picture depending on the 
cohomology of the resulting group (see the analogous discussion on homotopy in section
{\bf \ref{terms}}).

\vspace{3mm}
We point out in closing that the important question of investigating the situation at the quantum 
level was not answered in this note, and this certainly should be studied.

\bigskip\bigskip
\noindent
%%%%%%%%%%%%%%%%%%%%%%%%%%%%%%%%%%%%%%%%%%%%%%%%%%
{\bf \large Acknowledgements}\\
%%%%%%%%%%%%%%%%%%%%%%%%%%%%%%%%%%%%%%%%%%%%%%%%%%
\vspace{2mm}
\noindent
We thank Nitu Kitchloo for very useful discussions.
%\newpage

%%%%%%%%%%%%%%%%%%%%%%%%%%%%%%%%%%%%%%%%%%%%%%%%%%%%%%%%%%%%%%%%%%%%%


\begin{thebibliography}{99}
%%%%%%%%%%%%%%%%%%%%%%%%%%%%%%%%%%%%%%%%%%%%%%%%%%%%%%%%%%%%%%%%%%%%%
\bibitem{S5}
H. Sati, {$E_8$ gauge theory and gerbes in string theory},
[{\tt arXiv:hep-th/0608190}].

\bibitem{Mick}
J. Mickelsson,
{\it Gerbes, (twisted) K-theory, and the supersymmetric WZW model},
in Infinite dimensional groups and manifolds, IRMA Lect. Math. Theor. 
Phys. {\bf 5}, de Gruyter, Berlin, 2004,  
[{\tt arXiv:hep-th/0206139}].


\bibitem{Flux}
E.~Witten, {\it On flux quantization in M-Theory and the effective
action},
J. Geom. Phys. {\bf 22} (1997) 1-13,
[{\tt arXiv:hep-th/9609122}].

\bibitem{DMW}
E.~Diaconescu, G.~Moore and E.~Witten,
{\it $E_8$ gauge theory, and a derivation of K-Theory from M-Theory},
Adv. Theor. Math. Phys. {\bf 6} (2003) 1031,
[{\tt arXiv:hep-th/0005090}].

\bibitem{MS}
V. Mathai and H. Sati, {\it Some relations between twisted
K-theory and $E_8$ gauge theory}, J. High Energy Phys. {\bf 0403} (2004) 
016,
[{\tt arXiv:hep-th/0312033}].

\bibitem{AE}
A.~Adams and J.~Evslin,
{\it The loop group of $E_8$ and K-Theory from $11d$},
J. High Energy Phys. {\bf 02} (2003) 029,
[{\tt arXiv:hep-th/0203218}].

\bibitem{DFM}
E.~Diaconescu, D. Freed, and G.~Moore,
{\it The M-theory 3-form and $E_8$ gauge theory},
[{\tt arXiv:hep-th/0312069}].

\bibitem{BJSV}
M. Bershadsky, A. Johansen, V. Sadov, and C. Vafa,  
{\it Topological reduction of 4d SYM to 2d sigma models},
Nucl. Phys. {\bf B448} (1995) 166, 
[{\tt arXiv:hep-th/9501096}].

\bibitem{AB}
M. Atiyah and R. Bott,
{\it The Yang-Mills equations over Riemann surfaces}, 
Philos. Trans. Roy. Soc. London Ser. A {\bf 308} (1983) 523.

\bibitem{Kit}
N. Kitchloo,
{\it The Baum-Connes conjecture for loop groups}, Oberwolfach Reports
{\bf 27} (2005) 29.  

\bibitem{KM}
N. Kitchloo and J. Morava, 
{\it Thom prospectra for loop group representations},
in Elliptic Cohomology, London Mathematical Society Lecture Note Series {\bf 342},
Cambridge University Press, 2007.


\bibitem{Mein}
E. Meinrenken,
{\it The Basic gerbe over a compact simple Lie group}, 
Enseign. Math. {\bf (2) 49} (2003) 307, [{\tt arXiv:math.DG/0209194}]. 


\bibitem{ES}
J. Evslin and H. Sati, {\it SUSY vs $E_8$ gauge theory in 11
dimensions}, J. High Energy Phys. {\bf 0305} (2003) 048, [{\tt
arXiv:hep-th/0210090}].

\bibitem{Nish}
H. Nishino, {\it Supersymmetric Yang-Mills theory in eleven dimensions},
Phys. Lett. {\bf B 492} (2000) 201, [{\tt arXiv: hep-th/0008029}].

\bibitem{BL}
L. Baulieu and C. Laroche, 
{\it On Generalized self-duality equations towards supersymmetric 
quantum field theories as forms}, 
Modern Phys. Lett. {\bf A 13} (1998) 1115, [{\tt arXiv: hep-th/9801014}].

\bibitem{disk}
A. Sengupta, {\it Yang-Mills on surfaces with boundary: quantum theory and 
symplectic limit}, Comm. Math. Phys. {\bf 183} (1997) 661.

\bibitem{F95}
D. S. Freed, {\it Quantum groups from path integrals}, 
in Particles and Fields, Springer, New York, 1999, 
 [{\tt arXiv:q-alg/9501025}].

\bibitem{BV}
A. Bergman and U. Varadarajan,
{\it Loop groups, Kaluza-Klein reduction and M-theory},
J. High Energy Phys. {\bf 0506} (2005) 043,
[{\tt arXiv:hep-th/0406218}].

\bibitem{CS}
R. L. Cohen and A. Stacey,
{\it Fourier decompositions of loop bundles},
in  Homotopy theory: relations with algebraic geometry, group
cohomology, and algebraic $K$-theory,  85--95, Contemp. Math. {\bf 346}
AMS, Providence, RI, 2004, [{\tt arXiv:math.AT/0210351}].

\bibitem{Hit}
N. J. Hitchin,
{\it The self-duality equations on a Riemann surface}, 
Proc. London Math. Soc. {\bf (3) 55} (1987) 59. 

\bibitem{Hig}
M. Murray and D. Stevenson,
{\it Higgs fields, bundle gerbes and string structures},
Comm. Math. Phys. {\bf 243} (2003) 541,
[{\tt arXiv:math.DG/0106179}].

\bibitem{KY}
Y. Kuramoto and T. Yasui, 
{\it On Haefliger's obstructions to embeddings and transfer maps}, 
Osaka J. Math. {\bf 40} (2003) 69. 

\bibitem{Hae}
A. Haefliger, 
{\it Differentiable imbeddings}, 
Bull. Amer. Math. Soc. {\bf 67} (1961) 109. 

\bibitem{BT}
R. Bott and L. Tu, {\it Differential forms in algebraic topology},
Springer, NY, 1982. 

\bibitem{BCRS}
L. Bonora, P. Cotta-Ramusino, M. Rinaldi, and J. Stasheff, 
{\it The Evaluation map in field theory, sigma-models and strings II}, 
Comm. Math. Phys. {\bf 114} (1988) 381. 


\bibitem{avatars}
A. Losev, G. W. Moore, N. Nekrasov, and S. Shatashvili, 
{\it Four-dimensional avatars of two-dimensional RCFT},
Nucl. Phys. Proc. Suppl. {\bf 46} (1996) 130, 
[{\tt arXiv:hep-th/9509151}].

\bibitem{IKU}
T. Inami, H. Kanno, and T. Ueno, {\it Higher-dimensional WZW model on K\"ahler manifold and toroidal Lie algebra}, Mod. Phys. Lett. {\bf A12} (1997) 2757,
[{\tt arXiv:hep-th/9704010}].

\bibitem{MN}
G. Moore and P. Nelson, {\it  The Etiology of sigma model anomalies},
Comm. Math. Phys. {\bf 100} (1985) 83.





\end{thebibliography}
\end{document}